\documentclass[pra,twocolumn,superscriptaddress]{revtex4-2}
\usepackage{float,graphicx,multirow, amsmath,color,dsfont}
\usepackage{amsmath,bm}
\usepackage{enumerate}
\usepackage{amssymb,mathrsfs,dsfont}
\usepackage{amssymb,mathbbol}
\usepackage{amsfonts}
\usepackage{mathrsfs}
\usepackage{float}
\usepackage{xcolor,hyperref}
\definecolor{darkblue}{rgb}{0,0.0.1,0.3}
\definecolor{darkred}{rgb}{0.6,0.1,0}
\hypersetup{colorlinks,breaklinks,
	linkcolor=darkred,urlcolor=darkblue,
	anchorcolor=darkred,citecolor=
	darkred,pdfauthor=ChRiSh, pdftitle=ChRiSh.Thermal.Estimation}
\usepackage{tikz}
\usepackage{mathbbol}
\usepackage{float}
\usepackage[normalem]{ulem}

\begin{document}
	
	\title{Enhanced phase estimation in parity detection based Mach-Zehnder interferometer using non-Gaussian two-mode squeezed thermal input state }
	\author{Chandan Kumar}
	\email{chandan.quantum@gmail.com}
	\affiliation{Department of Physical Sciences,
		Indian
		Institute of Science Education and
		Research Mohali, Sector 81 SAS Nagar,
		Punjab 140306 India.}
	\author{Rishabh}
	\email{rishabh1@ucalgary.ca}
	\affiliation{Department of Physics and
		Astronomy, University of Calgary, Calgary T2N1N4, Alberta,
		Canada.}
	\author{Shikhar Arora}
	\email{shikhar.quantum@gmail.com}
	\affiliation{Department of Physical Sciences,
		Indian
		Institute of Science Education and
		Research Mohali, Sector 81 SAS Nagar,
		Punjab 140306 India.}	
	
	\begin{abstract}

While the quantum metrological advantages of performing non-Gaussian operations on two-mode squeezed vacuum (TMSV) states have been extensively explored, similar studies in the context of two-mode squeezed thermal (TMST) states are severely lacking.  In this paper, we explore the potential advantages of performing non-Gaussian operations on TMST state for phase estimation using parity detection based Mach-Zehnder interferometry. To this end, we consider the realistic model of photon subtraction, addition, and catalysis. We first provide a derivation of the unified Wigner function of the photon subtracted, photon added and photon catalyzed TMST state, which to the best of our knowledge is not available in the existing literature. This Wigner function is then used to obtain the expression for the phase sensitivity. Our results show that performing non-Gaussian operations on TMST states can enhance the phase sensitivity for significant ranges of squeezing and transmissivity parameters. We also observe that incremental advantage provided by performing these non-Gaussian operations on the TMST state is considerably higher than that of performing these operations on the TMSV state. Because of the probabilistic nature of these operations, it is of utmost importance to take their success probability into account. We identify the photon catalysis operation performed using a high transmissivity beam splitter as the optimal non-Gaussian operation when the success probability is taken into account. This is in contrast to the TMSV case, where we observe photon addition to be the most optimal. These results will be of high relevance for any future phase estimation experiments involving TMST states. Further, the derived Wigner function of the non-Gaussian TMST states will be useful for state characterization and its application in various quantum information protocols.

	\end{abstract}
	\maketitle
	\section{Introduction}
 
  Optical interferometers have played an important role in the advancement of quantum sensing~\cite{Dowling-cp-2008, Giovannetti2011}. In particular, Mach-Zehnder interferometer (MZI) has been extensively utilized for phase sensitivity studies. While the phase sensitivity of MZI using classical resources is limited by so called
	shot-noise limit (SNL)~\cite{caves-prd-1981}, quantum resources, such as, squeezed states and entangled states can break the SNL and achieve the Heisenberg limit (HL)~\cite{Giovannetti-science-2004,Hofmann-pra-2007, Anisimov-prl-2010,Jeong-prl-2019}. Many quantum states of light have been employed in parity measurement based optical interferometry to reach HL with the notable example of N00N states~\cite{gerry-pra-2000,gerry-pra-2001,gerry-pra-2002,gerry-pra-2003,gerry-pra-2010,gerry-cp-2010,Anisimov-prl-2010,ecs-prl-2011,Seshadreesan_2011,Plick_2010,aravind-2011,sesha-pra-2013,sahota-pra-2013,zhang-pra-2013}. However,  photon loss renders N00N states fragile to environmental interactions. Moreover, it has been shown that two-mode squeezed vacuum  (TMSV)    states can even break the HL limit~\cite{Anisimov-prl-2010}, but their implementation is limited by the maximum achievable squeezing~\cite{15dB}.

To overcome the hindrance caused by the upper limit on the maximum achievable squeezing, one can resort to the engineering of non-Gaussian states by performing photon subtraction (PS), photon addition (PA), or photon catalysis (PC) operations on the TMSV state. These non-Gaussian states have  enhanced squeezing and entanglement, which have been utilised for performance improvement in quantum  key distribution~\cite{qkd-pra-2013,qkd-pra-2018,qkd-pra-2019,qk2019,chandan-pra-2019,zubairy-pra-2020}, quantum teleportation~\cite{tel2000,tel2009,catalysis15,catalysis17,wang2015}
as well as   quantum metrology~\cite{gerryc-pra-2012,josab-2012,braun-pra-2014,josab-2016,pra-catalysis-2021,crs-ngtmsv-met}.

Thermal states are an important class of mixed Gaussian states,
	which have played a key role in quantum optics from its inception~\cite{hbt-1956}. These states have since then been used in various practical applications, such as, thermal lasers~\cite{thermal-lasers}, ghost imaging~\cite{gimaging2004,giamging2005,giamging2005a}, and quantum illumination~\cite{Illumination}. Two-mode squeezed thermal (TMST) state have been proposed for use in quantum phase estimation\cite{chinPB-thermal}. 
	TMST states have also been realized experimentally~\cite{tmst-prl-2014,tmst-prl-2016,tmst-2016,tmst-njp-2016}.
	Further, non-Gaussian operations on squeezed thermal states are also being considered actively. For instance, nonclassicality and entanglement  have been explored in photon subtracted and added squeezed thermal states~\cite{Hu-2010,Meng:JOSAB12,Hu:JOSAB12,Hu2:JOSAB12}. Similarly, photon subtracted TMST (PSTMST) and photon added TMST (PATMST) states have also been proposed as resources for quantum teleportation~\cite{Annalen-tele}.
	 There have been experimental efforts in the preparation, reconstruction and statistical parameter estimation of multiphoton subtracted thermal states~\cite{thermal-sub-2017,thermal-subtraction-2021}.
	
	Motivated by these studies, we  aim to determine whether non-Gaussian operations such as PS, PA and PC can improve the phase sensitivity of the original TMST state. 
	To this end, we consider the experimental model of PS, PA and PC operations~\cite{njp-2015} (see Fig.\ref{figsub}).  Next, we obtain the
 Wigner function describing the   PSTMST, PATMST and   photon catalyzed  TMST (PCTMST) states, which is utilized to evaluate the phase sensitivity of the parity detection based MZI. In this paper, we will collectively refer PSTMST, PATMST, and PCTMST states as NGTMST states, where ``NG" stands for non-Gaussian. Also note that from here on, the term non-Gaussian operations will be used only to refer PA, PS, and PC operations unless stated otherwise.

   We emphasize that the presence of an additional parameter in the TMST state  (average number of photons in the thermal
state) significantly enhances the complexity of analytical calculations of the involved quantities as compared to TMSV state. We employ the   phase space formalism rather than the more often used operator formalism because of the computational ease provided by the former~\cite{Fan_2003}.

  Our  results   clearly demonstrate  that  the non-Gaussian operations can significantly enhance a TMST state's phase sensitivity. 
To better understand the above results and to  find out the squeezing and transmissivity values rendering an enhancement in phase sensitivity using NGTMST resource states, we have shown the plots for the difference between the phase sensitivity of the TMST and NGTMST states in squeezing-transmissivity plane. Further, with a view to identify optimal non-Gaussian operations, we take the probabilistic nature of non-Gaussian operations into our consideration. A careful analysis shows that only for photon catalysis, the parameter values corresponding to large enhancement overlap with those corresponding to a region of high success probability, which happens in a large transmissivity regime. Hence, it can be concluded that of all non-Gaussian operations, implementing PC using a high transmissivity beam splitter is the optimal choice. This is contrary to the case involving non-Gaussian TMSV state, where PA operation comes out to be most optimal~\cite{crs-ngtmsv-met}.

  Also, the expression for  unified Wigner function of NGTMST states derived here, do not appear in the existing literature to  the best  of our knowledge. This  expression    will be  a welcome addition to existing literature and will find important use while  dealing with various CV QIP protocols involving NGTMST states.
  Additionally,
	 we supply a single expression for the parity detection-based phase sensitivity to deal with all three non-Gaussian operations performed on the TMST state at once.

	The layout of this paper is as follows.  In Sec.~\ref{cvsystem}, we present a brief overview of the formalism   for  CV systems. In Sec.~\ref{sec:wig}, we derive the   unified  Wigner function for   all  the NGTMST states. Section~\ref{sec:mzi} contains a short description of parity-detection based lossless MZI. We start by comparing the phase sensitivities of the TMSV and TMST states.   We  then show the advantages of using NGTMST states over TMST states to determine the phase for specific chosen values of involved parameters. Next, we move on to a more involved analysis by studying the difference between the phase sensitivity of the TMST and NGTMST states and the success probability of corresponding non-Gaussian operations for experimentally reasonable values of squeezing and transmissivity. To gain a better perspective, we also compare our results to the case of non-Gaussian TMSV state (Ref~\cite{crs-ngtmsv-met}).   Finally, in Sec.~\ref{sec:conc}, we conclude the article by summarizing the main points and discussing possible
future directions.
 
	\section{Brief description of CV systems}
	\label{cvsystem}
	
We consider an $n$-mode quantum optical system, whose $i^{\text{th}}$ mode can be represented by  a pair of Hermitian quadrature operators $\hat{q}_i$ and $\hat{p}_i$. For convenience, we define the following $2n$-dimensional column vector comprising of $n$ pairs of Hermitian quadrature operators: 
\begin{equation}\label{eq:columreal}
	\hat{ \xi} =(\hat{ \xi}_i)= (\hat{q}_{1},\,
	\hat{p}_{1} \dots, \hat{q}_{n}, 
	\, \hat{p}_{n})^{T}, \quad i = 1,2, \dots ,2n.
\end{equation}
In terms of the column vector $\hat{\xi}$, the canonical commutation relation can be formulated as
\begin{equation}\label{eq:ccr}
	[\hat{\xi}_i, \hat{\xi}_j] = i \Omega_{ij}, \quad (i,j=1,2,...,2n),
\end{equation}
where we have taken $\hbar$=1, and $\Omega$ is a 2$n$ $\times$ 2$n$ matrix given by
\begin{equation}
	\Omega = \bigoplus_{k=1}^{n}\omega =  \begin{pmatrix}
		\omega & & \\
		& \ddots& \\
		& & \omega
	\end{pmatrix}, \quad \omega = \begin{pmatrix}
		0& 1\\
		-1&0 
	\end{pmatrix}.
\end{equation}
The photon annihilation and creation operators for the $i^{\text{th}}$ mode are defined in terms of the corresponding quadrature operators as follows:
\begin{equation}
	\label{realtocom}
	\hat{a}_i=   \frac{1}{\sqrt{2}}(\hat{q}_i+i\hat{p}_i),
	\quad  \hat{a}^{\dagger}_i= \frac{1}{\sqrt{2}}(\hat{q}_i-i\hat{p}_i).
\end{equation}
For a quantum system with density operator $\hat{\rho}$, we can define the Wigner distribution function as below:
\begin{equation}\label{eq:wigreal}
	W(\bm{\xi}) = \int \frac{\mathrm{d}^n \bm{q'}}{{(2 \pi)}^{n}}\, \left\langle
	\bm{q}-\frac{1}{2}
	\bm{q}^{\prime}\right| \hat{\rho} \left|\bm{q}+\frac{1}{2}\bm{\bm{q}^{\prime}}
	\right\rangle \exp(i \bm{q^{\prime T}}\cdot \bm{p}),
\end{equation}
where
$\bm{\xi} = (q_{1}, p_{1},\dots, q_{n},p_{n})^{T} \in \mathbb{R}^{2n}$,
$\bm{q^{\prime}} \in \mathbb{R}^{n}$ 
and $\bm{q} = (q_1,
q_2, \dots, q_n)^T$, 
$\bm{p} = (p_1, p_2, \dots, p_n)^T $.
The Wigner  function of a Fock state $|n\rangle$ can be evaluated using Eq.~(\ref{eq:wigreal})   as
\begin{equation}\label{wig:fock}
	W_{|n\rangle}(q,p)=\frac{(-1)^n}{\pi}\exp  \left( 
	-q^2-p^2 \right)\,L_{n}\left[ 2(q^2+p^2) \right],
\end{equation}
where $L_n\{\bullet\}$ is the Laguerre polynomial of nth order.
We can also reformulate the Wigner function in terms of the
average of displaced parity operator~\cite{parity-1977}:
\begin{equation}\label{wigparity}
	W(\bm{\xi}) =\frac{1}{{ \pi}^{n}} \text{Tr} \left[ \hat{\rho}\, D(\bm{\xi}) \hat{\Pi} D^{\dagger} (\bm{\xi}) \right] ,
\end{equation}
where  $ \hat{\Pi} =\prod_{i=0}^{n}  \exp\left( i \pi   \hat{a}^{\dagger}_i \hat{a}_i \right)$
is the parity operator and $D(\bm{\xi}) = \exp[i \hat{ \xi} \, \Omega \, \bm{\xi}]$ is the displacement operator.

Gaussian states are an important class of CV-system states whose Wigner distribution is a Gaussian function.
The Wigner function~(\ref{eq:wigreal}) for $n$-mode Gaussian states 
  simplifies to the following form~\cite{weedbrook-rmp-2012}:
\begin{equation}\label{eq:wignercovariance}
	W(\bm{\xi}) = \frac{\exp[-(1/2)(\bm{\xi}-\bm{d})^TV^{-1}
		(\bm{\xi}-\bm{d})]}{(2 \pi)^n \sqrt{\text{det}V}},
\end{equation}
where  $\bm{d} =
	\text{Tr}[\hat{\rho} \,\hat{\xi}]$ is the displacement vector and $V$ is a $2n\times2n$ covariance matrix, whose element can be calculated as
\begin{equation}\label{eq:cov}
V_{ij}=\frac{1}{2}\langle \{\Delta \hat{\xi}_i,\Delta
	\hat{\xi}_j\} \rangle,
\end{equation}
where $\Delta \hat{\xi}_i = \hat{\xi}_i-\langle \hat{\xi}_i
\rangle$, and $\{\,, \, \}$ denotes anti-commutator.
The action of an infinite-dimensional unitary operator $\mathcal{U}$ on a density operator can be mapped to a symplectic transformation $S \in Sp(2n,\, \mathcal{R})$   acting on the quadrature operators. Further, 
 the state evolution $\rho \rightarrow \,\mathcal{U}  \rho
\,\mathcal{U}^{\dagger}$ can be rephrased as
a symplectic transformation in phase space as follows~\cite{arvind1995}:
\begin{equation}\label{transformation} 
	\bm{d}\rightarrow S \bm{d},\quad V\rightarrow SVS^T,\quad  \text{and} \,\, W(\xi) \rightarrow W(S^{-1} \xi).
\end{equation}

 	Quantum optical mode in thermal equilibrium with a bath at a given temperature is said to be in a thermal state. This mode can be thought of as a classical mixture of   different photon number states with weight factors given by the Boltzmann distribution.  As stated earlier in this paper, we consider the TMST state, which can be generated by subjecting two uncorrelated thermal modes to a two mode squeezing transformation.
	The TMST state is   described by the following covariance matrix:
	\begin{equation}
		V_{A_1A_2}=S_{A_1A_2}(r) V_{\text{th}} S_{A_1A_2}(r)^T, 
	\end{equation}
	where $V_{\text{th}} = (n_\text{th}+1/2)\mathbb{1}_4$ is the covariance matrix of the two uncorrelated
	thermal modes with $n_\text{th}$  being the average number of photons in the thermal
state. Further, $S_{A_1A_2}(r)$ is the two mode squeezing transformation given by
	\begin{equation}\label{tms}
		S_{A_1A_2}(r) = \begin{pmatrix}
			\cosh r \,\mathbb{1}_2& \sinh r \,\mathbb{Z} \\
			\sinh r \,\mathbb{Z}& \cosh r \,\mathbb{1}_2,
		\end{pmatrix}
		,\quad \mathbb{Z} =  \begin{pmatrix}
		1& 0 \\
		0& -1
	\end{pmatrix}.
	\end{equation}
Putting   $n_\text{th}=0$ in Eq.~(\ref{tms}), we obtain the covariance matrix of the TMSV state.
	Since TMST state is a Gaussian state with zero mean and covariance matrix given by Eq.~(\ref{tms}),
	the  Wigner function of the TMST state can be readily evaluated using Eq.~(\ref{eq:wignercovariance}):
	\begin{equation}
		\begin{aligned}
			W(\xi) =  \frac{1}{(2 \pi \kappa )^2} \exp\big[&- 
			(q_1^2+p_1^2+q_2^2+p_2^2)\cosh (2r)/(2\kappa) \\
			&+   (q_1 q_2-p_1 p_2) \sinh (2r)/\kappa	\big],
		\end{aligned}
	\end{equation}
	where $ \kappa =(n_\text{th}+1/2)$. 
	We now discuss PS, PA, and PC operations on a TMST state.

	
	\section{Wigner  function of NGTMST  states}\label{sec:wig}

\begin{figure}[h!]
	\includegraphics[scale=1]{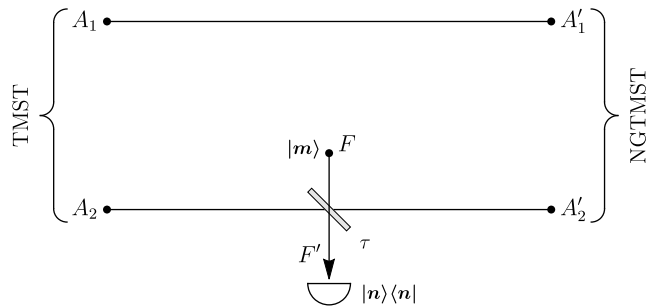}
	\caption{Schematic for the preparation of NGTMST states.  The mode $A_2$ of the TMST state
		is mixed  with  Fock state $|m\rangle$ using a beam
		splitter of transmissivity $\tau$. Photon number
		measurement  is performed on the output auxiliary mode to obtain the NGTMST states.}
	\label{figsub}
\end{figure}
  The schematic for the generation of NGTMST states is shown in Fig.~\ref{figsub}. 
Mode  $A_2$   of the TMST state is mixed with auxiliary mode $F$, which is in Fock state $|m\rangle$,  via beam-splitter  of transmissivity $\tau$.  
We represent the state of the three mode system before the beam splitter operation via its Wigner function:  
\begin{equation}
	W_{  A_1 A_2 F}(\xi) =  W_{A_1 A_2}(\xi_1,\xi_2) W_{|m\rangle}(\xi_3)  ,
\end{equation}
where $\xi_i=(q_i,p_i)^T$ $(i=1,2,3)$. The   beam-splitter    transforms the phase space variables $( \xi_2,\xi_3 )^T$ via the symplectic transofrmation
\begin{equation}\label{beamsplitter}
	B_{A_2 F}(\tau) = \begin{pmatrix}
		\sqrt{\tau} \,\mathbb{1}_2& \sqrt{1-\tau} \,\mathbb{1}_2 \\
		-\sqrt{1-\tau} \,\mathbb{1}_2&\sqrt{\tau} \,\mathbb{1}_2
	\end{pmatrix}.
\end{equation}
The  Wigner function after the beam-splitters operation is given by 
\begin{equation}
	\begin{aligned}
		W_{  A_1' A_2' F'}(\xi)   =W_{A_1 A_2 F}( [\mathbb{1}_2 \oplus B_{A_2 F}(\tau)]^{-1}\xi)  .
	\end{aligned}
\end{equation}
  A photon number resolving detector, represented by the positive-operator-valued measure (POVM) $\{\Pi_{n }=|n \rangle\langle n |,\mathbb{1}-\Pi_{n }\}$ is employed to measure the mode $F^{'}$. When   the POVM element  $\Pi_{n}$   clicks, the non-Gaussian operation is performed successfully on  the mode $A_2$. The expression for corresponding probability is provided in Eq.~(\ref{probeqq}). The reduced state is NGTMST states, and its 
  unnormalized Wigner   function is given by
\begin{equation}\label{detect}
	\begin{aligned}
		\widetilde{W}^{\text{NG}}_{A_1' A_2'}(\xi_1,\xi_2)=& 2 \pi\int  d^2 \xi_3  
		\underbrace{W_{A_1' A_2' F'}(\xi_1,\xi_2,\xi_3 )}_{\text{Three mode entangled state}}\\
		&\times 
		\underbrace{W_{|n \rangle }(\xi_3)}_{\text{Projection on }|n\rangle \langle n|} .\\
	\end{aligned}
\end{equation}
By carefully selecting the values of the parameters $(m, n)$, we can perform the required non-Gaussian operations. For PS operation, $m < n$, whereas for PA operation,     $m > n$. Lastly, for PC operation, $m = n$. PS operation  on TMST states yields PSTMST states. Similarly, PA and PC operations on TMST states yields PATMST and PCTMST states, respectively.

The generating function for the Laguerre polynomials
\begin{equation}\label{gen}
	\begin{aligned}
		L_n[2(q^2+p^2)]=	\bm{\widehat{D}}\exp \left[\frac{st}{2}+s(q+ip)-t(q-ip)\right],
	\end{aligned}
\end{equation}
with
\begin{equation}
	\bm{\widehat{D}} = \frac{2^n}{n!}  \frac{\partial^n}{\partial\,s^n} \frac{\partial^n}{\partial\,t^n} \{ \bullet \}_{s=t=0}.
\end{equation}
can be used to convert Eq.~(\ref{detect}) into a Gaussian integral. On integration of Eq.~(\ref{detect}), we get 
\begin{equation}\label{eq4}
	\begin{aligned}
	\widetilde{W}^{\text{NG}}_{A_1' A_2'}&=\frac{\exp \left(\bm{\xi}^T M_1 \bm{\xi}\right)}{a_0} \bm{\widehat{D}_1} \exp  \big[-a_1 u_1 v_1+a_2 u_1+a_3 v_1 \\
		& -a_4 u_2 v_2 +a_5 u_2+a_6 v_2+ a_7  (u_1 u_2+v_1 v_2 ) \big] ,\\
	\end{aligned}
\end{equation}
where the column vector $\bm{\xi}$ is defined as 
$(q_1,p_1,q_2,p_2)^T$ and the coefficients $a_i$ and the matrix $M_1$ are defined in Eqs.~(\ref{appwiga})  and  (\ref{appwigb}) of the Appendix~\ref{appsec}, respectively. 
Further, the differential operator $\bm{\widehat{D}_1} $ is given as
\begin{equation}
\begin{aligned}
	 \bm{\widehat{D}_1} = \frac{(-2)^{m+n}}{m!\,n!} \frac{\partial^{m}}{\partial\,u_1^{m}} \frac{\partial^{m}}{\partial\,v_1^{m}} 
	 \frac{\partial^{n}}{\partial\,u_2^{n}} \frac{\partial^{n}}{\partial\,v_2^{n}} \{ \bullet \}_{\substack{u_1= v_1=0 \\ u_2= v_2=0}}.\\
\end{aligned}
\end{equation}
We can express Eq.~(\ref{eq4}) in terms of two-variable 
Hermite polynomials $H_{m,n}(x,y)$:
\begin{equation}
	\begin{aligned}
		&\widetilde{W}^{\text{NG}}_{A_1' A_2'}=\frac{(-2)^{m+n}}{m!\,n!} \frac{\exp \left(\bm{\xi}^T M_1 \bm{\xi}\right)}{a_0} \sum_{i,j=0}^{\text{min}(m,n)}   \frac{a_1^{m}}{\sqrt{a_1}^{i+j}} \\
		&\times \frac{a_4^{n}}{\sqrt{a_4}^{i+j}} \frac{a_7^{i+j}}{i!\,j!}   \frac{m!}{(m-i)!} \frac{m!}{(m-j)!} \frac{n!}{(n-i)!} \frac{n!}{(n-j)!}\\
		&\times H_{m-i,m-j}\left[\frac{a_2}{\sqrt{a_1}},\frac{a_3}{\sqrt{a_1}}\right] 
		H_{n-i,n-j}\left[\frac{a_5}{\sqrt{a_4}},\frac{a_6}{\sqrt{a_4}}\right].
	\end{aligned}
\end{equation}
The probability  of a successful non-Gaussian operation can be evaluated as
\begin{equation}\label{probeqq}
	P^{\text{NG}}= \int d^2 \xi_1 d^2 \xi_2 \widetilde{W}^{\text{NG}}_{A_1' A_2'}
	=d_0 \bm{\widehat{D}_1}\exp \left(\bm{u}^T M_2 \bm{u}\right),
\end{equation}
where the column vector $\bm{u}$ is defined as 
$(u_1,v_1,u_2,v_2)^T$ and the matrix $M_2$ and coefficient $d_0$ are defined in Eqs.~(\ref{appprob1}) and~(\ref{appprob2})  of the
Appendix~\ref{appsec}.
The normalized  Wigner characteristic function $W^{\text{NG}}_{A'_1 A'_2}$ of the non-Gaussian NGTMST state is obtained as
\begin{equation}\label{normPS}
	W^{\text{NG}}_{A'_1 A'_2}(\xi_1,\xi_2) ={\left(P^{\text{NG}}\right)}^{-1}\widetilde{W}^{\text{NG}}_{A_1' A_2'}(\xi_1,\xi_2).
\end{equation}

 The Wigner function of several special cases can be readily derived from Eq.~(\ref{normPS}). 
For example, in the unit transmissivity limit $\tau  \rightarrow 1$   with  $m=0$, we obtain the Wigner function of the ideal PSTMST state $ \mathcal{N}_s \hat{a}_2^{n}  |\text{TMST}\rangle$ with $\mathcal{N}_s$ being the normalization factor. Similarly, in the unit transmissivity limit $\tau  \rightarrow 1$   with  $n=0$, we obtain the  Wigner function of the ideal PATMST state 
 $ \mathcal{N}_a \hat{a}{_2^{\dagger }}^{m}  |\text{TMST}\rangle$  with $\mathcal{N}_a$ being the normalization factor.
 Further, setting $ \kappa=1/2$ (equivalently $n_\text{th}=0$) in Eq.~(\ref{normPS}) yields the Wigner function of the non-Gaussian TMSV state.

	\section{Parity detection based phase sensitivity in MZI}\label{sec:mzi}
	
 \begin{figure}[h!]
 	\begin{center}
 		\includegraphics[scale=1]{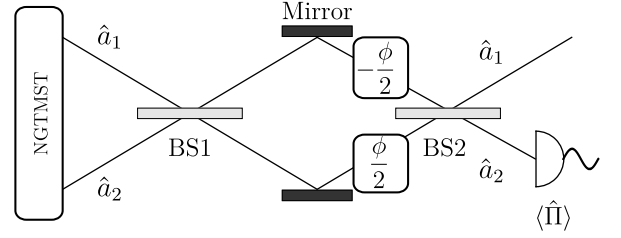}
 		\caption{   Schematic of     parity detection  based  Mach-Zehnder interferometer for   detection of  an unknown  phase shift  using NGTMST state as an input.}
 		\label{mzi}
 	\end{center}
 \end{figure}
 
We consider a lossless MZI comprised of two $50{:}50$ beam splitters and two   phase shifters, as shown in Fig.~\ref{mzi}.
The NGTMST states are introduced as  input   in  the interferometer. The annihilation operators $\hat{a}_1$ and $\hat{a}_2$ represent the two input modes. 
This setup is used for estimating the unknown phase $\phi$ introduced via the two phase shifters by measuring  the  parity operator on the output mode $\hat{a}_2$.

  To describe the mode transformations induced by the optical elements in the MZI,  it is convenient to use the Schwinger representation of $\text{SU}(2)$ algebra~\cite{yurke-1986}. 
The generators of the $\text{SU}(2)$ algebra, also known as angular momentum operators can be written in terms of the annihilation and creation operators of the input modes as follows:
\begin{equation}
	\begin{aligned}
		\hat{J}_1 = &\frac{1}{2}(\hat{a}^\dagger_1\hat{a}_2+\hat{a}_1\hat{a}^\dagger_2),\\
		\hat{J}_2 = &\frac{1}{2i}(\hat{a}^\dagger_1\hat{a}_2-\hat{a}_1\hat{a}^\dagger_2),\\
		\hat{J}_3 = &\frac{1}{2}(\hat{a}^\dagger_1\hat{a}_1-\hat{a}^\dagger_2\hat{a}_2).
	\end{aligned}
\end{equation}
These angular momentum operators follow the commutation relations $[J_i,J_j] = i \epsilon_{ijk}J_k $. The infinite-dimensional unitary transformations of the first and the second  beam splitters are given by $e^{-i(\pi/2)J_1}$ and $e^{i(\pi/2)J_1}$, respectively. Further, the cumulative action of the two phase shifters is given by $e^{i \phi J_3}$. 
Hence, the overall action of the lossless MZI is given by
\begin{equation}
	\mathcal{U}(S_{\text{MZI}}) = e^{-i(\pi/2)J_1}e^{i \phi J_3}e^{i(\pi/2)J_1}=e^{-i \phi J_2}.
\end{equation}
The resultant symplectic transformation $S_{\text{MZI}}$ of the MZI acting on the phase space variables $(\xi_1,\xi_2)^T$ is given by
\begin{equation} 
	S_{\text{MZI}} =  \begin{pmatrix}
		\cos (\phi/2) \,\mathbb{1}& -	\sin (\phi/2) \,\mathbb{1} \\
		\sin (\phi/2) \,\mathbb{1}& \cos (\phi/2) \,\mathbb{1}
	\end{pmatrix}.
\end{equation}
The transformation of the Wigner function of the input state due to the MZI  can be written using Eq.~(\ref{transformation}) as
\begin{equation}
	W_{\text{in}}(\xi ) \rightarrow W_{\text{in}}(S_{\text{MZI}}^{-1}\xi) =W_{\text{out} } (\xi).
\end{equation}
As shown in the schematic of MZI (Fig.~\ref{mzi}), we perform   parity measurement on the  output mode $\hat{a}_2$, which basically differentiates between   even and odd photons numbers Fock state. The corresponding  photon number parity operator  is written  as
\begin{equation}
	\hat{\Pi}_{\hat{a}_2} =  \exp\left( i \pi   \hat{a}^{\dagger}_2 \hat{a}_2 \right)=  (-1)^{\hat{a}_2^{\dagger}\hat{a}_2}.
\end{equation}
 Therefore, the average value of the parity measurement operator can be evaluated using  Eq.~(\ref{wigparity})~\cite{Birrittella-2021}:
\begin{equation}
	\langle \hat{\Pi}_{\hat{a}_2} \rangle =f(\phi)  = \pi \int \, d^2\xi_1 \, W_{\text{out}} (\xi_1,0).
\end{equation}
Using the Wigner function of the  NGTMST state~(\ref{normPS}), the average of the parity measurement operator comes out to be
\begin{equation}\label{apari}       
	f(\phi) = \frac{e_0 \bm{\widehat{D}_1}\exp \left(\bm{u}^T M_3 \bm{u}\right)}{d_0 \bm{\widehat{D}_1}\exp \left(\bm{u}^T M_2 \bm{u}\right)},
\end{equation}
where the matrix $M_3$ and coefficient $e_0$ are defined in Eqs.~(\ref{appparity1}) and~(\ref{appparity2})  of the
Appendix~\ref{appsec}.
Taking the unit transmissivity limit under the condition $m=n$ in Eq.~(\ref{apari}), yields the average of the parity operator for the case of input TMST state:
\begin{equation}\label{aparitmst}
	f(\phi)_\text{TMST} = \frac{1-\lambda^2}{2 \kappa [1+\lambda^4-2 \lambda^2 \cos(2 \phi)]^{1/2}}.
\end{equation}
Further, setting   $ \kappa=1/2$  in Eq.~(\ref{aparitmst}) renders the average of the parity operator for an input TMSV state. 
\par
The error propagation formula allows us to write the  phase uncertainty or sensitivity   as
\begin{equation}
	\Delta \phi = \frac{\sqrt{1-f(\phi+\pi/2) ^2}}{|\partial  f(\phi+\pi/2) /\partial \phi|}.
\end{equation}
 The phase uncertainty for the input TMST state can be written using Eq.~(\ref{aparitmst}) as
 \begin{equation}\label{phasetmst}
 		\Delta \phi_\text{TMST} = \displaystyle \frac{\sqrt{\kappa^2-\displaystyle\frac{(1-\lambda^2)^2}{1+\lambda^4-2 \lambda^2 \cos(2 \phi)}}}
 		{\bigg|\displaystyle\frac{\lambda^2(1-\lambda^2) \sin(2\phi)}{[1+\lambda^4-2 \lambda^2 \cos(2 \phi)]^{3/2}}\bigg|}.
 \end{equation}
 
We note that the phase uncertainty depends on the following parameters:
 \begin{enumerate}[(i)]
     \item Squeezing of the TMST state, referred to as squeezing from here on.
     \item Transmissivity of the beam splitter used in the implementation of the non-Gaussian operations, referred to as transmissivity from now onward. 
     \item Magnitude of unknown phase introduced, which will be referred to as phase.
     \item Average number of photons in the thermal state.
 \end{enumerate}

 \begin{figure}[h!]
	\begin{center}
		\includegraphics[scale=1]{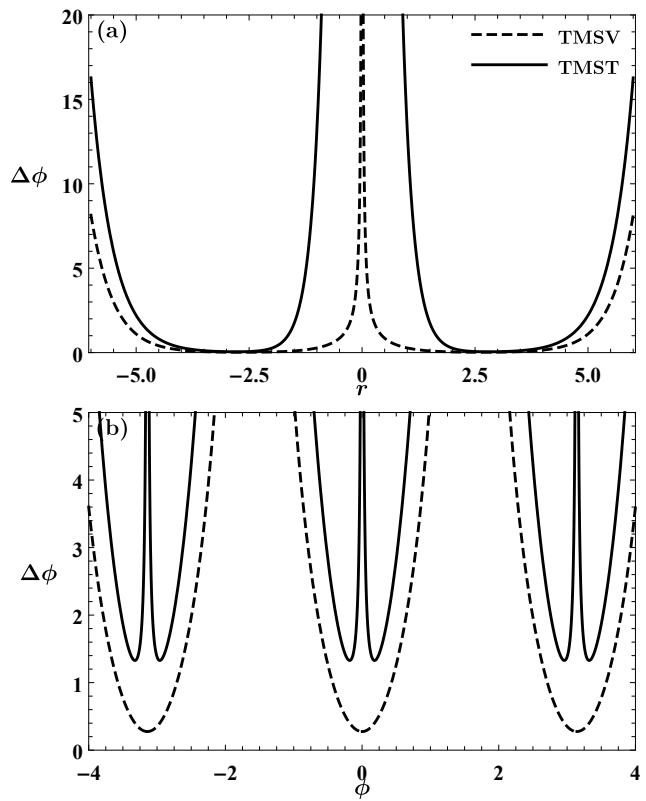}
		\caption{(a) Phase uncertainty $\Delta \phi $ as a function of squeezing parameter $r$ for the TMST and TMSV states.  We have taken $\phi=0.01$ rad. (b) Phase uncertainty $\Delta \phi $ as a function of phase $\phi$. We have taken $r=1$. Further, we have set $\kappa=1$ for the TMST state for both the panels.   While $\Delta \phi $ and $\phi$ axes are in rad, $r$ axis is dimensionless.  }
		\label{tmstparity}
	\end{center}
\end{figure}

 We now numerically study the dependence of the phase uncertainty of the input TMSV and TMST states on squeezing and  phase (the thermal parameter $\kappa$ has been set equal to 1 for TMST state throughout the article). The results are shown in Fig.~\ref{tmstparity}. 
 As is evident from Eq.~(\ref{phasetmst}), the phase uncertainty for the input TMST state will be larger than the input TMSV state, which can also seen in Fig.~\ref{tmstparity}. As is shown in Fig.~\ref{tmstparity}(a), the phase uncertainty blows up at $r=0$ for both the TMSV and TMST states and attains a minima at $r {\approx 2.65}$ and $r {\approx} 2.80$, respectively. However, such large values of squeezing cannot be achieved with current technology~\cite{15dB}. We note here that these specific numerical values of squeezing are for $\phi= 0.01$. 
 
 As is obvious from Eq.~(\ref{phasetmst}), the phase uncertainty varies periodically with period $\pi$ as a function of  phase,  which can also be seen in Fig.~\ref{tmstparity}(b). The phase uncertainty for the TMSV state is minimized at $\phi=0, \pm \pi, \pm 2\pi, \cdots$, whereas for the TMST state, it blows up at these values with a pair of minima appearing in the surrounding regions.

\subsection{Advantages of non-Gaussian operations on TMST state in phase estimation}
\label{subsec:tf}
\begin{figure}[h!]
	\begin{center}
		\includegraphics[scale=1]{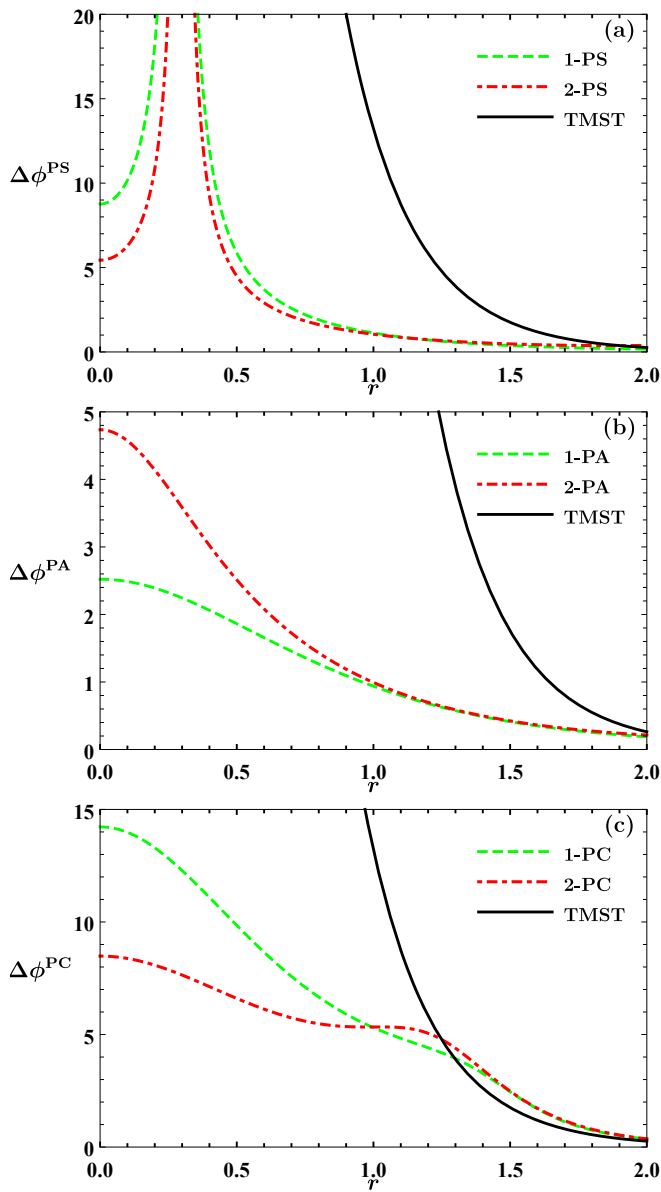}
		\caption{Phase uncertainty $\Delta \phi $   as a function of the  squeezing parameter $r$ for TMST and various NGTMST states.  We have set the $\tau=0.9$ and $\phi=0.01$ rad for all the panels. While $\Delta \phi $   axis is in rad, $r$ axis is dimensionless. }
		\label{phase_1d_sq}
	\end{center}
\end{figure} 
In this section, we explore the advantages of performing the non-Gaussian operations on TMST state in the context of phase estimation. As we shall see, various NGTMST states outperform the original TMST state. We first analyze the dependency of the phase uncertainty on initial squeezing ($r$) while transmissivity ($\tau$), and magnitude of phase ($\phi$) are kept fixed at suitably chosen values (see Fig.~\ref{phase_1d_sq}). In Fig.~\ref{phase_1d_sq}(a), we observe that 1-PSTMST and 2-PSTMST states improve the phase sensitivity over the original TMST states. It is also interesting to note that the value of $\Delta \phi^\text{PS}$ blows up at $r {\approx} 0.3$. Similar features can also be observed for other NGTMST states if parameter values are appropriately chosen.
As seen in Fig.~\ref{phase_1d_sq}(b), both  the 1-PATMST and 2-PATMST states outperform the original TMST state. We also notice that for upto $r{\approx} 1$, 1-PATMST state outperforms 2-PATMST state, which is in contrast with Fig.~\ref{phase_1d_sq}(a), where 2-PSTMST state outperforms 1-PSTMST state.
Figure~\ref{phase_1d_sq}(c) shows that 1-PCTMST and 2-PCTMST states outperform the TMST state up to  $r{\approx} 1.3$. While 1-PCTMST state performs better for $r {\lesssim} 1.0$, it is outperformed by 2-PCTMST state beyond this squeezing.

\begin{figure}[h!]
	\begin{center}
		\includegraphics[scale=1]{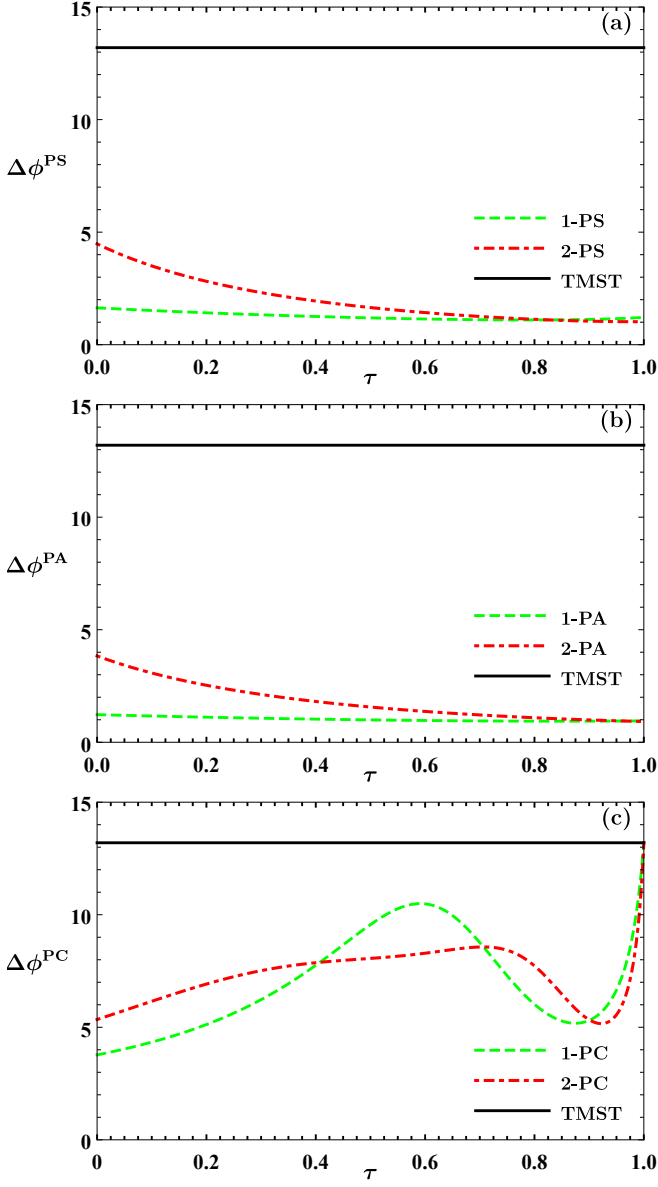}
		\caption{ Phase uncertainty $\Delta \phi $   as a function of the transmissivity $\tau$ for TMST and various NGTMST states.  We have set the squeezing   as $r=1$ for all the panels.  Further, we have taken $\phi=0.01$ rad in all the panels. While $\Delta \phi $   axis is in rad, $\tau$ axis is dimensionless.}
		\label{phase_1d_tr}
	\end{center}
\end{figure}

In Fig.~\ref{phase_1d_tr}, we show the
phase uncertainty with respect to transmissivity at a given squeezing and phase value. It can be clearly seen that for given values of squeezing and phase, all NGTMST states outperform the TMST state for all transmissivity values. We also notice that the phase sensitivity improves with increasing $\tau$ for PSTMST and PATMST states. Moreover, as we can see in Fig.~\ref{phase_1d_tr}(c), either the 1-PCTMST or 2-PCTMST states can provide better phase sensitivity depending on the value of transmissivity.  Also, the phase uncertainty of the PCTMST states approaches that of the TMST state in  the unit  transmissivity limit.

 \begin{figure}[h!]
 	\begin{center}
 		\includegraphics[scale=1]{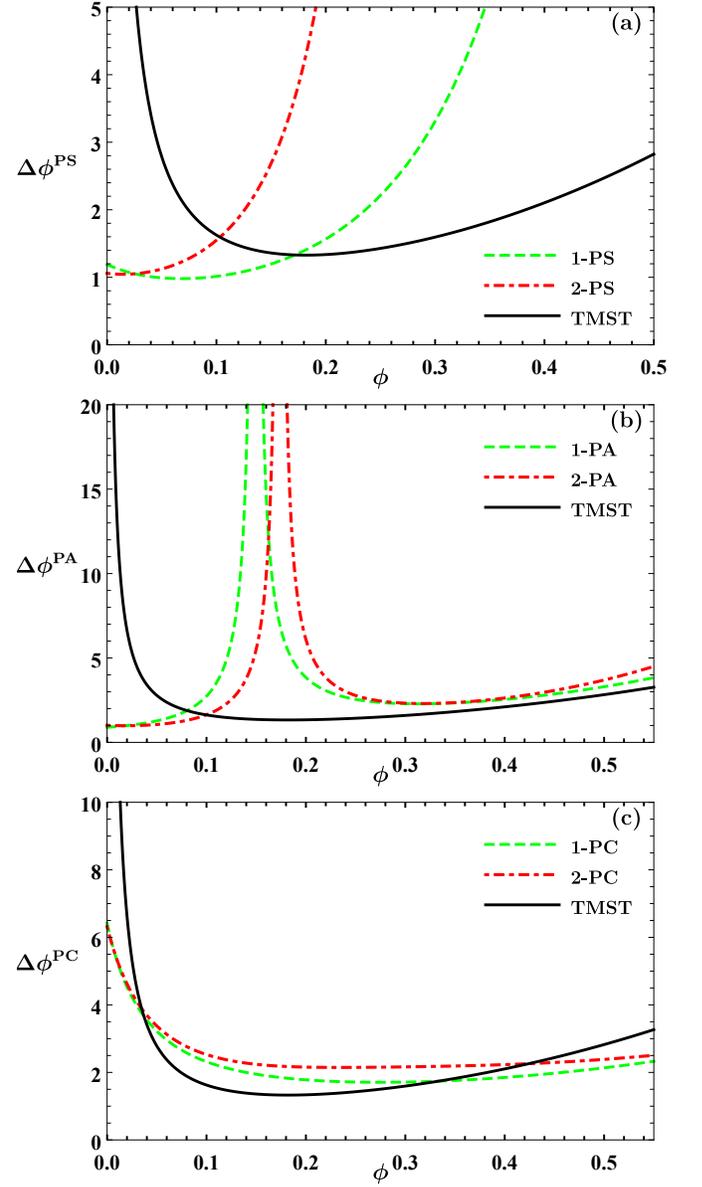}
 		\caption{ Phase uncertainty $\Delta \phi $   as a function of the phase $\phi$ for TMST and various NGTMST states.  We have set the $\tau=0.9$ and  $r=1$   for all the panels.    Both $\Delta \phi $ and $  \phi $   axes are in rad.}
 		\label{phase_1d_ph}
 	\end{center}
 \end{figure}
In Fig.~\ref{phase_1d_ph}, we plot the phase uncertainty with respect to phase magnitude at a given squeezing and transmissivity values.  While PSTMST and PATMST states improve the phase sensitivity for small phase values, PCTMST states can enhance the phase sensitivity even for larger phase values.

\subsection{Success probability and relative performances of NGTMST states in phase estimation}
 
In the previous section, we explored the advantages of using NGTMST states over the original TMST state for certain specifically chosen values of state parameters (~$r$ and $\tau$)  and phase  $\phi$.  In order to   gain a  comprehensive insight    into the comparative performances of NGTMST and TMST states in the context of phase estimation, we now explore the advantages of using non-Gaussian resources for a reasonable range of transmissivity and squeezing parameters for a given value of phase. For this purpose, we start by defining a figure of merit, $\mathcal{D}^{\text{NG} }$, as the difference of $\Delta\phi$ between TMST and NGTMST states:  
\begin{equation}\label{merit}
\mathcal{D}_{T}^{\text{NG} }= \Delta\phi^\text{TMST}- \Delta\phi^\text{NGTMST},
\end{equation}
where the subscript $T$ stands for thermal in the TMST state.
The region of positive $\mathcal{D}_T^{\text{NG} }$ corresponds to   those values of  transmissivity and squeezing  for which the NGTMST states perform better than the TMST state.

  To better understand the impact of the probabilistic nature of non-Gaussian operations, we plot the success probability alongside the $\mathcal{D}_T^{\text{NG} }$ for the same state parameter ranges. Success probability is also a good measure of resource utilization and can be defined as the fraction of successful non-Gaussian operations.

 \begin{figure}[h!]
	\begin{center}
		\includegraphics[scale=1]{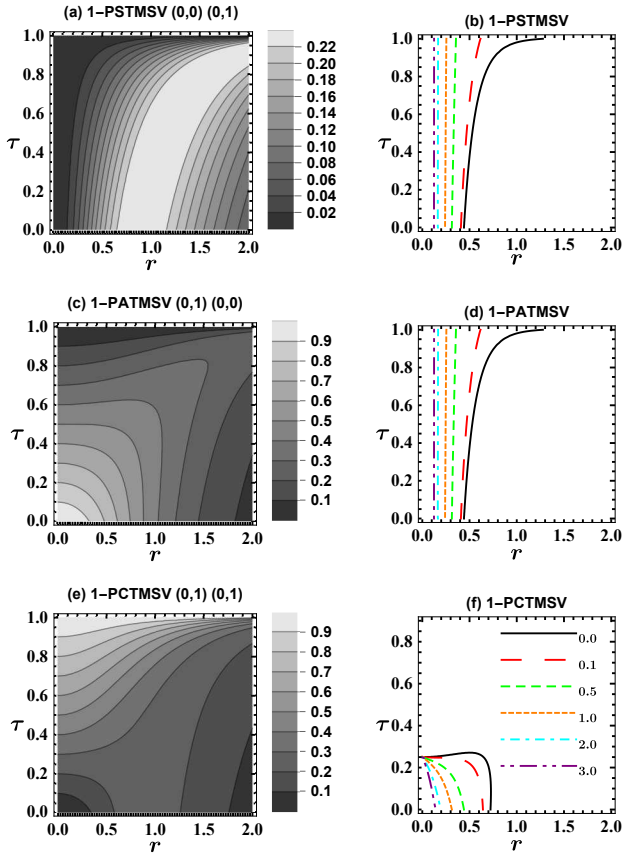}
		\caption{  Left panels show the success probability as a function of the transmissivity $\tau$ and
			squeezing parameter $\lambda$. Right panels show the curves of  fixed $ \mathcal{D}_{V}^{\text{NG} }$, the difference of $\Delta \phi$ between TMSV and NGTMSV states,  as a function of $\tau$ and $\lambda$. We have considered NGTMSV states generated by performing non-Gaussian operations on
			one of the modes of the TMSV state, and the corresponding values of the parameters $(m_1,n_1)(m_2,n_2)$ have been shown. The legend on the right panel shows the plotted values of $ \mathcal{D}_{V}^{\text{NG} }$.  The phase, $\phi$, is taken to be $0.01$.    Both the  axes are dimensionless. 
		 }
		\label{asymngtmsv}
	\end{center}
\end{figure}

 For comparative purposes, we reproduce the results of Ref.~\cite{crs-ngtmsv-met}   in Fig.~\ref{asymngtmsv}, which demonstrates the advantages of performing  non-Gaussian operations on the TMSV state in the context of phase estimation. The equivalent figure of merit~(\ref{merit}) for NGTMSV states can be written as
\begin{equation}
    \mathcal{D}_{V}^{\text{NG} }= \Delta\phi^\text{TMSV}- \Delta\phi^\text{NGTMSV},
\end{equation} 
where the subscript $V$ stands for vacuum in the TMSV state.
In the left panels of Fig.~\ref{asymngtmsv}, success probability for various non-Gaussian operations are plotted, whereas the right panels show the plot for different fixed values of corresponding  $\mathcal{D}_{V}^{\text{NG} }$~$(=0.0,\,0.1,\,0.5,\,1,\,2,\,3)$ as a function of the transmissivity $\tau$ and squeezing parameter $r$. 
Positive $\mathcal{D}_{V}^{\text{NG} }$ corresponds to that region in the $(\tau,r)$ space where the NGTMSV states outperform the TMSV state.  A careful comparison of different non-Gaussian operations shows that only for photon addition, the region of positive $\mathcal{D}_{V}^{\text{NG} }$ overlaps with the corresponding region of high success probability. Therefore, it can be concluded that out of all the non-Gaussian operations considered, only photon addition offers an advantage while taking the probabilistic nature of these operations into account. Reference~\cite{crs-ngtmsv-met} provides a detailed discussion of these results.

\begin{figure}[h!]
	\begin{center}
		\includegraphics[scale=1]{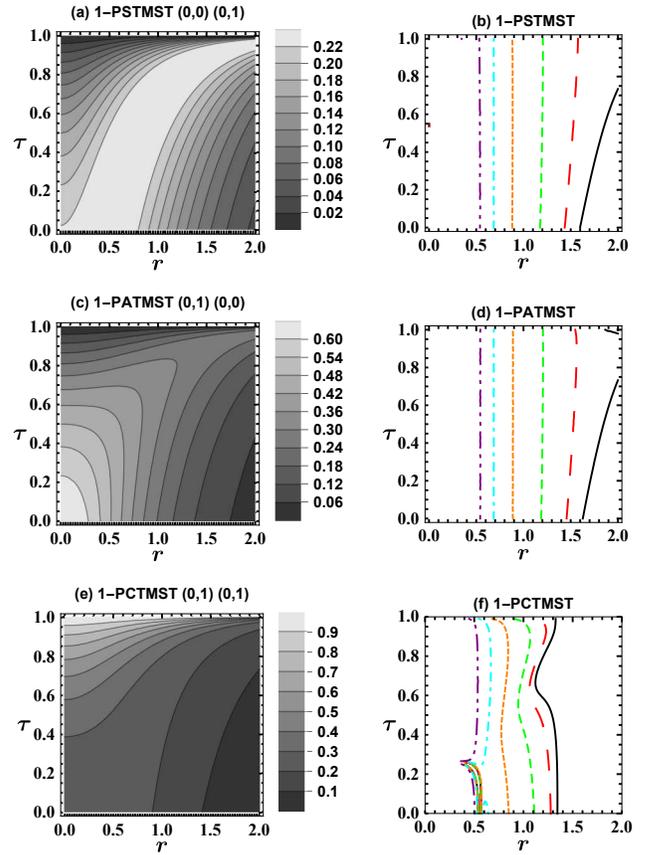}
		\caption{  Left panels show the success probability as a function of the transmissivity $\tau$ and
			squeezing parameter $\lambda$. Right panels show the curves of  fixed $ \mathcal{D}_{T}^{\text{NG} }$, the difference of $\Delta \phi$ between TMST and NGTMST states,  as a function of $\tau$ and $r$. We have considered NGTMST states generated by performing non-Gaussian operations on
			one of the modes of the TMST state, and the corresponding values of the parameters $(m_1,n_1)(m_2,n_2)$ have been shown.   The phase, $\phi$, is taken to be $0.01$.  Solid black, large dashed red, dashed green, dotted orange, dot dashed cyan, and double dot dashed purple curves represent  $ \mathcal{D}_{T}^{\text{NG} }$~$(=0,\,1,\,5,\,20,\,50,\,100)$, respectively.    Both the  axes are dimensionless. }
		\label{asymngtmst}
	\end{center}
\end{figure}

We now proceed to analyze the advantages rendered by the NGTMST states as compared to the TMST state.
As shown in Fig.~\ref{asymngtmsv}, we plot the success probability for various non-Gaussian operations in the left panel of Fig.~\ref{asymngtmst}, whereas the right panel show the plot for different fixed values of corresponding  $ \mathcal{D}_{T}^{\text{NG} }$~$(=0,\,1,\,5,\,20,\,50,\,100)$ as a function of the transmissivity $\tau$ and squeezing parameter $r$.  There is a considerable enhancement in the magnitude of $ \mathcal{D}_{T}^{\text{NG} }$ in comparison to $ \mathcal{D}_{V}^{\text{NG} }$. This  signifies the fact that   incremental advantage of performing    non-Gaussian operations on the TMST state   is much more  as compared to   that of performing these operations on the TMSV state.  

On careful comparison of the left panels of Figs.~\ref{asymngtmsv} and \ref{asymngtmst}, we observe that the maximum achievable success probability for PS and PC operations on the TMST state is approximately the same as on the TMSV state, whereas it decreases considerably for PA operation. Similarly, a careful comparison of the right panels reveals that the region of positive $\mathcal{D_{T}^{\text{NG}}}$ in $r$ and $\tau$ space increases for all three non-Gaussian operations. While for PC operation, this change is due to an increase in the allowed values of both parameters $r$ and $\tau$, for PA and PS operations, this is not the case as there is no scope for increment in $\tau$ and only the allowed range of $r$ is increased. 

Taking the success probability considerations into account, we observe that only for photon catalysis, the region of positive $\mathcal{D}_{T}^{\text{NG} }$ overlaps with the corresponding region of high success probability ( large $\tau$ regime). Therefore, photon catalysis offers maximum advantage while taking the probabilistic nature of non-Gaussian operations into account.


	\section{Conclusion}
	\label{sec:conc}
	In this article, we showed that performing non-Gaussian operations on TMST states enhances the phase sensitivity in parity detection based MZI. To this end, we  derived a unified Wigner   distribution  function describing PSTMST, PATMST, and PCTMST states altogether. We utilize this function to obtain  a single phase sensitivity expression for all three   cases mentioned above. By appropriately changing the  number of input   and detected photon, we can perform PS, PA and PC operations.   The Wigner function and the phase sensitivity depend on the initial   squeezing of the TMST state, the average number of photons in the thermal state,   the transmissivity of the beam splitter used in the implementation of the non-Gaussian operations, and the magnitude of the unknown phase introduced. 
	  A careful analysis involving the probabilistic nature of non-Gaussian operations reveals  the photon catalysis operation as the most optimal non-Gaussian operation.  These optimal conditions are achieved while working in a high transmissivity regime.

 It is clear that the results of this study will be of significant relevance for any future phase estimation experiments involving TMST states. Recent experiments implementing PS operations on thermal states signal that our proposal could be implemented in lab~\cite{thermal-subtraction-2017}. Further, the derived Wigner function will be of great use in the state characterization via quantifying nonclassicality~\cite{nc-2020}, entanglement~\cite{ent-prl-2010}, non-Gaussianity~\cite{Ivan2012,ng-2021}, and nonlocality~\cite{nonlocality}. As mentioned earlier, we overcame the calculational challenges involved in dealing with non-Gaussian operations on TMST state by following phase space formalism instead of the traditional operator method. The phase space method utilized in this work can be useful in circumventing calculations challenges in various other problems as well, for instance, quantum teleportation via PCTMST states, a problem posed in Ref.~\cite{Annalen-tele}. While the scope of this paper is limited to exploring the advantages of performing non-Gaussian operation on a single mode of the TMST state,   the effects of performing these non-Gaussian operations on both modes remain to be investigated.
	
	\section*{Acknowledgement}
	This is the third article in a publication series
	written in the celebration of the completion of 15 years of IISER Mohali. C.K.  acknowledges the financial
	support from {\bf DST/ICPS/QuST/Theme-1/2019/General} Project
	number {\sf Q-68}.

	\appendix

\section{Coefficients appearing in the Wigner function, probability and average of parity operator}\label{appsec}
The coefficients $a_i$ appearing in Eq.~(\ref{eq4}) are given as
\begin{equation}\label{appwiga}
\begin{aligned}
    a_0=&\pi ^2 k \left(2 k \mu ^2 T+\Lambda r^2\right)\mu ^{-2},\\
    a_1=&k r^2 \left(2 k \mu ^2+\Lambda\right)b_0^{-1},\\
    a_2=&(2b_1 \left(q_1-i p_1\right)+b_2 \left(q_2+i p_2\right))b_0^{-1},\\
    a_3=&-(2b_1 \left(q_1+i p_1\right)-i b_2 \left(p_2+i q_2\right))b_0^{-1},\\
    a_4=&-k r^2 \left(2 k \mu ^2-\Lambda\right)b_0^{-1},\\
    a_5=&(2b_3 \left(q_1-i p_1\right)+b_4 \left(q_2+i p_2\right))b_0^{-1},\\
    a_6=&-(2b_3 \left(q_1+i p_1\right)-i b_4 \left(p_2+i q_2\right))b_0^{-1},\\
    a_7=&4 k^2 \mu ^2 t b_0^{-1},\\
\end{aligned}
\end{equation}
where $\mu=\sqrt{1-\lambda^2}$, $t=\sqrt{\tau}$, $r=\sqrt{1-\tau}$, $T=(1+t^2)$, $\Lambda=(1+\lambda^2)$ and
\begin{equation}
\begin{aligned}
    b_0=&-2 k \left(2 k \mu ^2 T+\Lambda r^2\right), \\
    b_1=&2 k \lambda  r t,\\
    b_2=&-k r \left(2 k \mu^2+\Lambda\right),\\
    b_3=&2 k \lambda  r,\\
    b_4=&k r t \left(2 k \mu ^2-\Lambda\right).\\
\end{aligned}
\end{equation}
Further the matrix $M_1$ appearing in Eq.~(\ref{eq4}) is given as
\begin{equation}\label{appwigb}
    M_1=\frac{1}{b_0}\left(
\begin{array}{cccc}
 c_1 & 0 & c_2 & 0 \\
 0 & c_1 & 0 & -c_2 \\
 c_2 & 0 & c_3 & 0 \\
 0 & -c_2 & 0 & c_3 \\
\end{array}
\right),
\end{equation}
where
\begin{equation}
\begin{aligned}
    c_1=& k \Lambda T+\mu ^2 r^2,\\
    c_2=& -4 k \lambda  t, \\
    c_3=& k\Lambda T+2 k^2 \mu ^2 r^2.\\
\end{aligned}
\end{equation}
The explicit form of matrix $M_2$ appearing in Eq.~(\ref{probeqq}) is given as
\begin{equation}\label{appprob1}
\begin{aligned}
   M_2=\frac{1}{d_4} \left(
\begin{array}{cccc}
 0 & d_1 & 0 & d_2 \\
 d_1 & 0 & d_2 & 0 \\
 0 & d_2 & 0 & d_3 \\
 d_2 & 0 & d_3 & 0 \\
\end{array}
\right),
\end{aligned}
\end{equation}
where
\begin{equation}\label{appprob2}
\begin{aligned}
    d_0=&\frac{2 \mu ^2}{2 k \Lambda r^2+\mu ^2 T},\\
    d_1=&-r^2 \left(\mu^2+2 k \Lambda\right),\\
    d_2=&-2 \mu ^2 t,\\
    d_3=&r^2 \left(\mu ^2-2 k \Lambda\right),\\
    d_4=&8 k \Lambda r^2-2 \mu ^2 T.\\
\end{aligned}
\end{equation}

The explicit form of matrix $M_3$ appearing in Eq.~(\ref{apari}) is given as
\begin{equation}\label{appparity1}
    M_3=\frac{1}{e_7}\left(
\begin{array}{cccc}
 e_1 & e_2 & e_3 & e_4 \\
 e_2 & e_1 & e_4 & e_3 \\
 e_3 & e_4 & e_5 & e_6 \\
 e_4 & e_3 & e_6 & e_5 \\
\end{array}
\right),
\end{equation}
where
\begin{equation}\label{appparity2}
\begin{aligned}
    e_0=& 8\mu^2 e_7^{-1/2},\\
    e_1=&8 k \lambda  r^2 t \left(s_2 f_4-2s_1 f_1 \mu ^2\right),\\
   e_2=& f_6 r^2 \left(f_1 \mu ^2-c_1 f_4\right) ,\\
   e_3=& 8 k \lambda  r^2 s_1 \left(4 k \Lambda  r^2-f_3 \mu ^2 T\right) ,\\
   e_4=& -8 k t \left[\Lambda  \mu ^2 r^2 (c_1 f_1+f_2)+4 k T (\Lambda ^2-4 \lambda ^2 s_1^2)\right] ,\\
   e_5=& 8 k \lambda  r^2 t \left(s_2 f_5-2s_1 f_1 \mu ^2\right) ,\\
   e_6=& f_6 r^2 \left(f_1 \mu ^2-c_1 f_5\right) ,\\
    e_7=& 4 \left[\left(4 k \Lambda  T-f_3 \mu ^2 r^2\right)^2-(16 k \lambda s_1 t)^2\right],\\
\end{aligned}
\end{equation}
where $c_1{=}\cos  \, \phi $, $s_1{=}\sin \, \phi $, $c_2{=}\cos (2\phi)$, and $s_2{=}\sin (2\phi)$ and
\begin{equation}
\begin{aligned}
    f_1=& 4k^2-1 ,& \quad f_4=& f_2 \mu ^2+4 k \Lambda ,\\
    f_2=& 4k^2+1 ,& f_5=& f_2 \mu ^2-4 k \Lambda ,\\
    f_3=& c_1 f_1-f_2 ,& f_6=& f_4 \mu ^2 r^2-4 k \Lambda  T .\\
\end{aligned}
\end{equation}

 	\hspace{.1 in}


%

\end{document}